\documentclass[twocolumn,showpacs,preprintnumbers,amsmath,amssymb,superscriptaddress]{revtex4-1}

\usepackage{graphicx}
\usepackage{dcolumn}
\usepackage{bm}

\begin{document}

\title{Efficiency at maximum power of thermally coupled heat engines}

\author{Y. Apertet}\email{yann.apertet@u-psud.fr}
\affiliation{Institut d'Electronique Fondamentale, Universit\'e Paris-Sud, CNRS, UMR 8622, F-91405 Orsay, France}
\author{H. Ouerdane}
\affiliation{CNRT Mat\'eriaux UMS CNRS 3318, 6 Boulevard Mar\'echal Juin, F-14050 Caen Cedex, France}
\author{C. Goupil}
\affiliation{Laboratoire CRISMAT, UMR 6508 CNRS, ENSICAEN et Universit\'e de Caen Basse Normandie, 6 Boulevard Mar\'echal Juin, F-14050 Caen, France}
\author{Ph. Lecoeur}
\affiliation{Institut d'Electronique Fondamentale, Universit\'e Paris-Sud, CNRS, UMR 8622, F-91405 Orsay, France}

\date{\today}

\begin{abstract}
We study the efficiency at maximum power of two coupled heat engines, using thermoelectric generators (TEGs) as engines. Assuming that the heat and electric charge fluxes in the TEGs are strongly coupled, we simulate numerically the dependence of the behavior of the global system on the electrical load resistance of each generator in order to obtain the working condition that permits maximization of the output power. It turns out that this condition is not unique. We derive a simple analytic expression giving the relation between the electrical load resistance of each generator permitting output power maximization. We then focuse on the efficiency at maximum power (EMP) of the whole system to demonstrate that the Curzon-Ahlborn efficiency may not always be recovered: the EMP varies with the specific working conditions of each generator but remains in the range predicted by irreversible thermodynamics theory. We finally discuss our results in light of non-ideal Carnot engine behavior.
\end{abstract}


\maketitle

\section{\label{sec:intro}Introduction}

One of the  pillars of thermodynamics is the existence of a limitation in the conversion of heat into work: the second law, as expressed by Carnot, gives the \emph{maximum} efficiency that an ideal heat engine working between two temperature reservoirs $T_{\rm hot}$ and $T_{\rm cold}$ (with $T_{\rm hot} > T_{\rm cold}$) may boast. This maximum, known as the Carnot efficiency, here denoted $\eta_{\rm C}$, is only function of these temperatures: $\eta_{\rm C} = (T_{\rm hot} - T_{\rm cold}) / T_{\rm hot}$ \cite{Carnot1824}. This limit may be reached only if the process of conversion is fully reversible; but the ideal Carnot engine is a \emph{zero-power} engine since a reversible transformation in a thermodynamic system is \emph{quasi-static}: it requires an infinite time to complete. We can even go further by asserting that the Carnot engine is, as a matter of fact, unphysical: no dissipative element ensures its causality, so no arrow of time can be defined for such system.

>From a theoretical as well as a practical viewpoint, only models of irreversible heat engines must be considered to address the central points of causality and system evolution. The main consequence of the introduction of irreversibilities in models is the dependence of the engine's efficiency, $\eta$, and produced power, $P$, on the working conditions. This dependence ties $\eta$ and $P$ together: $\eta \equiv \eta(P)$. The Carnot efficiency can then be reached only when the engine works infinitely slowly; conversely, as the speed of the thermodynamic cycle of the engine increases, its conversion efficiency decreases. For very fast cycles the efficiency even reduces to zero. As regards power, if the cycle time $\tau$ is very long, it vanishes as it is given by the ratio of work, finite in that case, to the time $\tau$. On the contrary, if the speed increases too much, irreversibilities become preponderant and power vanishes again. Hence, between these two extremal cases, there are a variety of working conditions also leading to power maximization. For practical purposes, real thermodynamic engines must produce power, not just work: this observation shifted interest from efficiency, at the center of Carnot's derivations, to output power, thus establishing the theory of finite-time thermodynamics \cite{FTT1,FTT2,FTT3}; for a recent review, see Ref. \cite{Andresen2011}.

This theory, which emerged during the 1970s, may be viewed as a generalization of the thermodynamics of irreversible processes. The seeds were sown in the 1950s when physicists and engineers were working to obtain the best possible performances from the newly built atomic power plants: their design not only had to ensure power maximization but also an efficiency as high as possible for this particular working condition. To reach this purpose, a new model of heat engine was put forward: the irreversibilities were introduced by considering finite linear thermal conductances between heat reservoirs and the ideal heat engine, i.e. with no internal dissipations. This model was later referred to as \emph{endoreversible} system for irreversibilities occur outside the engine \cite{Rubin1979}. Basing his analysis of efficiency at maximum power [$\eta(P_{\rm max})$] on this then novel model, Novikov \cite{Novikov1957, Novikov1958} was the first to demonstrate the following seminal expression for EMP:

\begin{equation}
\eta(P_{\rm max}) = 1 - \sqrt{\frac{T_{\rm cold}}{T_{\rm hot}}},
\label{eq:CA}
\end{equation}

\noindent for endoreversible engines. Note that Novikov \cite{Novikov1957} attributes the above expression to Yvon \cite{Yvon1955} (however he points out that no demonstration was given). Nowadays, this result, very popular owing to its simplicity, is known as the Curzon-Ahlborn (CA) efficiency $\eta_{\rm CA}$, after the names of the two authors who rederived Eq.~(\ref{eq:CA}) independently from a refined analysis of a model heat engine in 1975 \cite{Curzon1975}. It is interesting to note that this expression was derived in 1957 by Chambadal \cite{Chambadal1957}, who thus shares the paternity of Eq.~(\ref{eq:CA}); however the hypothesis made on the heat exchange between the heat reservoirs and the engine are completely different from the ones made by Novikov, and Curzon and Ahlborn who assumed the linearity of thermal contacts. Hence, even if the final result is the same, one should keep in mind that the models are different. The CA efficiency turned out to be in good agreement with real EMP values, particularly in the case of power plants \cite{Curzon1975}. It has then rapidly been considered as a universal upper bound for EMP of heat engines \cite{Gordon89, Leff87}.

In 2005 Van den Broeck stated that the CA efficiency derives from the theory of linear irreversible thermodynamics, and that it can be generalized to any heat engine working in the regime of strong flux coupling \cite{VandenBroeck2005}; one of the main interests of this article is the elements that it brings to the debates in finite-time thermodynamics. First, Van den Broeck used the Onsager's reciprocal relations of the force-flux formalism\cite{Onsager1,*Onsager2,Callen1,Domenicali1954} to study heat engines. Finite-time thermodynamics analyses thus moved from cyclic to steady-state engine operations. The cycle time is no longer a parameter to optimize in that case: power maximization is obtained by tuning the thermodynamic forces applied to the engine. One should note that the EMP and all features associated to heat engines with cyclic operations remain the same for steady-state operations. Second, Van den Broeck highlighted the fact that the CA efficiency can only be obtained for engines operating in the \emph{strong coupling} regime. This assumption implies that inside the heat engine, the heat flux and the matter flux, which yield the output power, are tightly linked; more precisely they are proportional in the frame of linear theory.

The use of Onsager coefficients allows a natural and transparent description of the internal dissipations and thus puts forward the question of the influence of irreversibilities on EMP. Indeed, there are mainly three types of dissipation sources for heat engines: friction, heat leaks and finiteness of the heat transfert rate between the heat reservoirs and the engine. For a heat engine to produce power, the working fluid, which transports heat, has to move; such displacement of working fluid generates a heat flux proportionnal to the matter flux. Hence, any additionnal exchange of heat between temperature reservoirs can only result in a loss: energy flows through the system without any additional production of power. As discussed in Ref.~\cite{Apertet2011c}, heat leaks cannot be used to ensure causality: they exist independently of the working fluid displacement and, as such, they may be considered as a parallel parasitic process. For endoreversible models, causality is due to finiteness of the heat transfert rate. Friction as source of irreversibility and its consequences on EMF are discussed in detail in Sec.~\ref{sec:discussion}.

To demonstrate that the CA efficiency is, as a matter of fact, an upper bound for EMP, reached only when strong coupling is imposed, Van den Broeck used a cascade construction of heat engines in series \cite{VandenBroeck2005}. Each generator of this chain placed between two heat reservoirs at temperatures $T_{\rm hot}$ and $T_{\rm cold}$ respectively, is assumed to be working at maximum output power; this type of setup was further investigated by Jim\'enez de Cisneros and Calvo Hern\'andez \cite{Jimenez2007, Jimenez2008}. These previous works considered generic heat engines; in the present article, we focuse on a particular type of system: thermally coupled thermoelectric generators. As emphasized by de Groot, \emph{``The phenomena of thermoelectricity have always served as touchstones for various theories of irreversible phenomena''}\cite{deGroot}.

In a recent work \cite{Apertet2011c}, our choice to use a TEG as a model system to examine the impact of irreversibilities sources on EMP, proved fruitful. We demonstrated that the CA efficiency is specific to endoreversible engines and hence cannot be used to characterize engines with internal dissipation. The fact that our results contradict the main message of Van den Broeck in Ref.~\cite{VandenBroeck2005}, i.e. the universality of the CA efficiency for engines working at maximum power in the strong coupling regime, provided the impetus to investigate the behavior of a chain of thermoelectric generators in the strong coupling regime in order to uncover and explain the subtleties at the heart of the physics of energy conversion efficiency of engines operating at maximum output power.

Contrary to the works presented in Refs.~\cite{VandenBroeck2005, Jimenez2007, Jimenez2008} we do not consider an infinite chain but rather a simple (though nontrivial, as it turns out) cascade construction made of only two heat engines. Our article is organized as follows. In Sec.~\ref{sec:simu} we use a numerical model to study the power and efficiency of the whole system depending on the particular electrical load for each TEG. In Sec.~\ref{sec:analytic}, using the concept of effective thermal conductance introduced in Ref.~\cite{Apertet2012b}, we derive an analytical expression for both maximum power and efficiency at maximum power. We go further in the analysis of the EMP using our numerical model in Sec.~\ref{sec:EMP} where we consider cases when the two TEGs characteristics are very different in order to study the impact of dissipation localization. Sec.~\ref{sec:intdiss} is devoted to internal dissipation. A discussion of our findings followed by concluding remarks end the article. Further detail on strong flux coupling and Joule heating are provided in a series of appendices.

\section{\label{sec:simu}Simulation of two TEG thermally coupled in series}

The study of a system composed of two thermoelectric modules thermally coupled in series is sufficient to provide insight into the behavior and main characteristics of a cascade construction of coupled heat engines.

\subsection{Framework}

\begin{figure}
	\centering
		\includegraphics[width=0.5\textwidth]{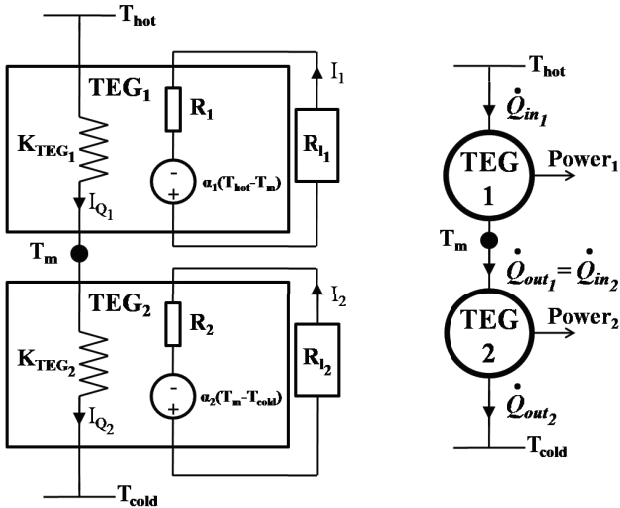}
	\caption{Thermoelectric (left) and thermodynamic pictures of the thermoelectric generators thermally in series.}
	\label{fig:figure1}
\end{figure}

Before considering the whole system, we give the phenomenological relations that define the behavior of each generators. The individual behavior of each TEG, placed between two temperature reservoirs at $T_{\rm hot}$ and $T_{\rm cold}$, can be described using a force-flux formalism, relating the electrical and thermal currents, $I$ and $I_Q$ to the electrical potential and temperature differences, $\Delta V$ and $\Delta T$, across the generator \cite{Apertet2012b}:

\begin{equation}\label{frcflx}
\left(
\begin{array}{c}
I\\
I_Q\\
\end{array}
\right)
=
\frac{1}{R}~
\left(
\begin{array}{cc}
1~ & ~\alpha\\
\alpha \overline{T}~ & ~\alpha^2 \overline{T} + RK_{_{I=0}}\\
\end{array}
\right)
\left(
\begin{array}{c}
\Delta V\\
\Delta T\\
\end{array}
\right),
\end{equation}

\noindent Each TEG is characterized by its isothermal electrical resistance $R$, its thermal conductance at zero electrical current (open circuit)$K_{_{I=0}}$, and its thermopower (or Seebeck coefficient) $\alpha$. The quantity $\overline{T}$ is the average temperature inside the TEG, which we take as $\overline{T} = (T_{\rm hot} + T_{\rm cold})/2$. The generator performances are characterized by the so-called figure of merit $ZT$ defined by $ZT = \alpha \overline{T}/(RK_{_{I=0}})$.

Heat is transported by conduction inside the TEG but advection also is a process to account for in presence of a nonzero electrical current \cite{Apertet2012b} since transport of heat directly is related with the motion of charge carriers in the particular direction set by the flow of electric charges in the system. Both heat transport processes may be associated into an equivalent thermal conductance $K_{\rm TEG}$ for the generator that includes advective and conductive terms. In the present work we consider TEGs in the strong flux coupling regime: the electrical current and the thermal current are proportionnal. This condition implies that the thermal conductance at zero electrical current is zero and that $K_{\rm TEG}$ reduces to the advective part of the thermal conductance(a more exhaustive explanation is given in Appendix A). This assumption is necessary to reach the Carnot efficiency and recover the Curzon-Ahlborn value at maximum power for some specific working conditions \cite{VandenBroeck2005}.

Consider the association of two TEGs thermally in series as shown in Fig.~\ref{fig:figure1}. The quantities defined above, now associated with each generator, are denoted with subscripts $1$ and $2$ respectively. One of the main hypotheses we make for this composite model system is the adiabaticity of the connection between the two TEGs: all the heat released from TEG$_1$ is entirely injected in TEG$_2$; there is no heat loss at the junction. This relation of continuity is the key point to derive the intermediate temperature, $T_{\rm m}$, at the junction point between the two TEGs.

The heat flux rejected by TEG$_1$ reads:

\begin{equation}
I_{Q_1}^{\rm (out)} = \alpha_1 T_{\rm m} I_1 + \frac{R_1 I_1^2}{2}
\label{eq:IQout1}
\end{equation}

\noindent Application of Ohm's law yields the electrical current in TEG$_1$: $I_1 = \alpha_1 (T_{\rm hot} - T_{\rm m}) / (R_1 + {R_{\ell}}_1)$, so we obtain :

\begin{equation}
I_{Q_1}^{\rm (out)} = \frac{\alpha_1^2 T_{\rm m} (T_{\rm hot}-T_{\rm m})}{R_1 + {R_{\ell}}_1} + \frac{R_1}{2} \left[\frac{\alpha_1 (T_{\rm hot}-T_{\rm m})}{R_1 + {R_{\ell}}_1}\right]^2.
\label{eq:IQ1out}
\end{equation}

\noindent The additional term associated with Joule heating is discussed in the Appendix C. Similarly, the incoming heat flux inside TEG$_2$ is given by:

\begin{equation}
I_{Q_2}^{\rm (in)} = \frac{\alpha_2^2 T_{\rm m} (T_{\rm m}-T_{\rm cold})}{R_2 + {R_{\ell}}_2} + \frac{R_2}{2} \left[\frac{\alpha_2 (T_{\rm m}-T_{\rm cold})}{R_2 + {R_{\ell}}_2}\right]^2.
\label{eq:IQ2in}
\end{equation}

\noindent Continuity of heat current, $I_{Q_1}^{\rm (out)}=I_{Q_2}^{\rm (in)}$, yields a second-order equation in $T_{\rm m}$ (which we solve numerically for a large range of values of the load electric resistances ${R_{\ell}}_1$ and ${R_{\ell}}_2$). The knowledge of the intermediate temperature naturally leads to the values of both power and efficiency for the whole system as well as for each TEG:

\begin{subequations}
\begin{equation}
P_1 = \frac{{R_{\ell}}_1 \alpha_1^2 (T_{\rm hot}-T_{\rm m})^2}{({R_{\ell}}_1 + R_1)^2}
\label{eq:power1}
\end{equation}
\noindent and
\begin{equation}
P_2 = \frac{{R_{\ell}}_2 \alpha_2^2 (T_{\rm m}-T_{\rm cold})^2}{({R_{\ell}}_2 + R_2)^2}
\label{eq:power2}
\end{equation}
\end{subequations}

The total output power $P$ simply is the sum of $P_1$ and $P_2$. The conversion efficiency is defined by :

\begin{equation}
\eta = \frac{P}{I_{Q}^{\rm (in)}}
\label{eq:eff}
\end{equation}

\noindent where $I_{Q}^{\rm (in)}$ is given by:

\begin{equation}
I_{Q}^{\rm (in)} \equiv I_{Q_1}^{\rm (in)} =  \frac{\alpha_1^2 T_{\rm hot} (T_{\rm hot}-T_{\rm m})}{R_1 + {R_{\ell}}_1} + \frac{R_1}{2} \left[\frac{\alpha_1 (T_{\rm hot}-T_{\rm m})}{R_1 + {R_{\ell}}_1}\right]^2
\label{eq:IQin}
\end{equation}

\subsection{Results}

\begin{figure}
	\centering
		\includegraphics[width=0.5\textwidth]{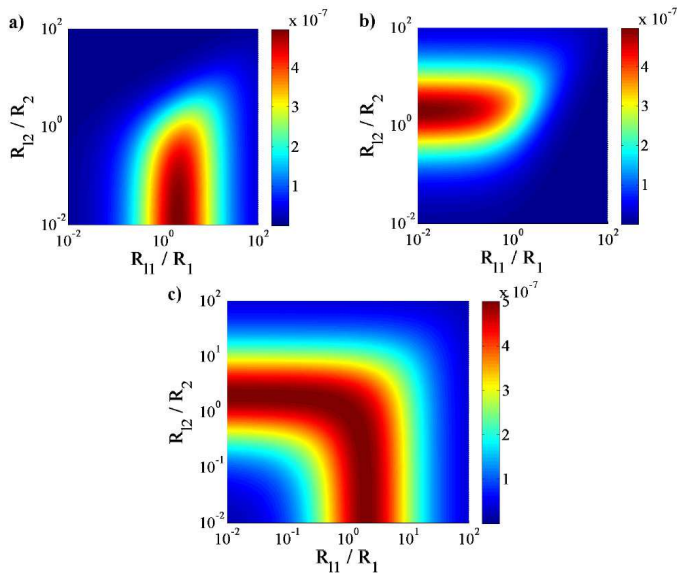}
	\caption{Output powers as functions of the load resistances scaled to the internal resistances ${R_{\ell}}_1/R_1$ and ${R_{\ell}}_2/R_2$: panel a for TEG$_1$; panel b for TEG$_2$; panel c for the whole system.}
	\label{fig:figure2}
\end{figure}

\begin{figure}
	\centering
		\includegraphics[width=0.50\textwidth]{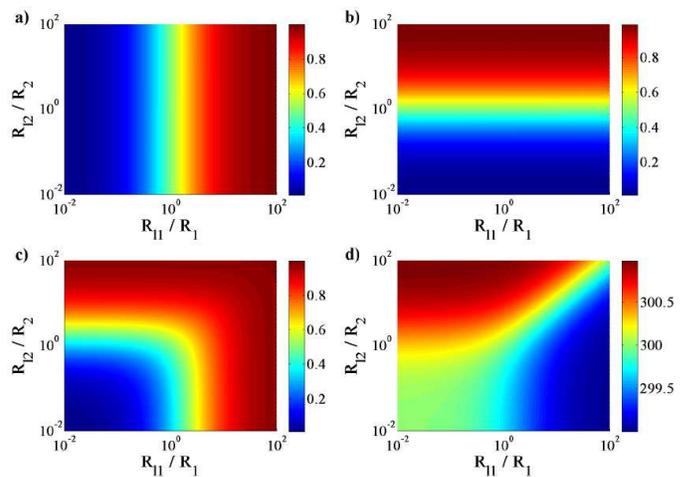}
	\caption{Conversion efficiencies (scaled to the Carnot efficiency) as functions of the load resistances ${R_{\ell}}_1/R_1$ and ${R_{\ell}}_2/R_2$: panel a for TEG$_1$; panel b for TEG$_2$; panel c for the whole system. Intermediate temperature $T_{\rm m}$ on panel d.}
	\label{fig:figure3}
\end{figure}

The electrical output powers produced and the related efficiencies (scaled to the Carnot efficiency, $\eta_{\rm C}$) for each thermogenerator and for the whole system as functions of the load resistances ${R_{\ell}}_1$ and ${R_{\ell}}_2$, are plotted in Figs.~(\ref{fig:figure2}) and (\ref{fig:figure3}). For these simulations we took fixed values for the temperatures: $T_{\rm hot} = $ 301 K and $T_{\rm cold}=$ 299 K; the TEGs are similar: $R_1 = R_2 = 0.01 \Omega$, and $\alpha_1 = \alpha_2 = 0.1$ mV K$^{-1}$. For the sake of clarity the load resistances are normalized to the coresponding TEG internal resistance.

A first striking result is the appearence of a whole set of couples of load resistances $({R_{\ell}}_1,{R_{\ell}}_2)$ leading to the maximum power: each couple corresponds to a particular working condition. Another interesting result is that the two generators never work at maximum power at the same time. We even observe the existence of an intermediate range within which both generators are far from optimal performances. However the maximum power of the whole system seems to remain constant even in these intermediate working conditions. Hence, for a chain of heat engines operating in the strong coupling regime, there is no direct relation between the particular working condition of each component and the global working condition; this confirms the findings presented in Ref.~\cite{Jimenez2008}.

To analyze the efficiency of the chain we use an expression obtained first by Harman \cite{Harman1958}. For $N$ heat engines thermally coupled in series, he demonstrated that the total efficiency $\eta$ can be expressed as a function of the efficiency $\eta_i$ of each engine composing the chain:

\begin{equation}
\eta=1-\prod_{i=1}^N {(1-\eta_i)}.
\label{eq:harman}
\end{equation}

\noindent The above expression reduces to:

\begin{equation}
\eta=\eta_1+\eta_2-\eta_1 \eta_2
\label{eq:harman2}
\end{equation}

\noindent for two heat engines.

For each generator the plots showing the efficiency as function of working conditions can be simply explained: for a specific generator, efficiency, \emph{unlike} power, depends only upon the associated load resistance and not upon the other generator working conditions. So, if $R_{\ell} \gg R$, i.e. for an open circuit, the efficiency tends to the Carnot value (under the  assumption of strong coupling) whereas if $R_{\ell} \ll R$, i.e. for a short circuit, the efficiency tends to zero. From there, the global efficiency trend can be obtained from the Harman relation, Eq.~(\ref{eq:harman}), which we use to also check that, as derived by Van den Broeck \cite{VandenBroeck2005}, the efficiency defined by $\eta=1-\left(\frac{T_{\rm cold}}{T_{\rm hot}}\right)^a$ reproduces itself upon concatenation of heat engines. Equation~(\ref{eq:harman}) also shows that if an engine works with zero efficiency, it does not modify the efficiency of the whole system: thus a chain composed of two engines with one operating at zero efficiency and the other one at the Carnot efficiency, works with an efficiency equal to the Carnot value. One does not need all the engines of the chain to work at maximum efficiency to make the chain work at maximum efficiency, as stated in \cite{Jimenez2008}.

\section{\label{sec:analytic}Analytic derivations}

To gain further insight into the behavior of the coupled TEGs, we derive simple analytical expressions for output power and efficiency at maximum power.

\subsection{Derivation of the intermediate temperature $T_{\rm m}$}

The calculations that follow are based on the assumption of small conversion efficiency, which is easily satisfied in the linear regime. Adopting the same approach as in Ref. \cite{Apertet2012b}, we obtain:
 
\begin{equation}
T_{\rm m} = \frac{K_{\rm TEG_{1}}T_{\rm hot} + K_{\rm TEG_{2}}T_{\rm cold}}{K_{\rm TEG_{2}} + K_{\rm TEG_{1}}}
\end{equation}

\noindent with

\begin{subequations}
\begin{equation}
\label{eq:effthermalcond1}
K_{\rm TEG_{1}}= \frac{\alpha_1^2 (T_{\rm hot}+T_{\rm m})/2}{R_1+{R_{\ell}}_1}
\end{equation}
\begin{equation}
\label{eq:effthermalcond2}
K_{\rm TEG_{2}}= \frac{\alpha_2^2 (T_{\rm m}+T_{\rm cold})/2}{R_2+{R_{\ell}}_2}
\end{equation}
\end{subequations}

\noindent By making the reasonnable approximation that $ \overline{T} \approx(T_{\rm m}+T_{\rm cold})/2 \approx (T_{\rm hot}+T_{\rm m})/2$, we end up with the following analytical formula for $T_{\rm m}$:

\begin{equation}
T_{\rm m} \approx  \frac{\alpha_1^2/(R_1+{R_{\ell}}_1) T_{\rm hot} + \alpha_2^2/(R_2+{R_{\ell}}_2) T_{\rm cold}}{\alpha_1^2/(R_1+{R_{\ell}}_1) + \alpha_2^2/(R_2+{R_{\ell}}_2)}
\label{eq:Tm}
\end{equation}

\noindent Equation~(\ref{eq:Tm}) is useful to better understand Fig.~(\ref{fig:figure3}.d), and the asymptotic values of the curves, in particular. When both TEGs are shorted (which amounts to having ${R_{\ell}}_1 \ll R_1$ and ${R_{\ell}}_2 \ll R_2$), the temperature $T_{\rm m} $ is the weigthed average of the reservoir temperatures; each weight being the power factor of the related TEG. If only one TEG is shorted, $T_{\rm m}$ becomes close to the associated reservoir temperature. For example, when only TEG$_1$ is shorted, ${R_{\ell}}_1 \ll R_1$ and ${R_{\ell}}_2 \gg R_2$, the temperature $T_{\rm m}$ tends to $T_{\rm hot}$.

\subsection{Output power}

Knowledge of the analytical form \eqref{eq:Tm} of the temperature $T_{\rm m} $ permits to rewrite the individual output powers of the two thermogenerators given in Eqs.~(\ref{eq:power1}) and (\ref{eq:power2}); the ouput power for TEG$_1$ now is given by:

\begin{equation}
P_1 = \alpha_1^2 (T_{\rm hot}-T_{\rm cold})^2 \frac{{R_{\ell}}_1 \alpha_2^4}{[({R_{\ell}}_1 + R_1) \alpha_2^2 + ({R_{\ell}}_2 + R_2) \alpha_1^2]^2}
\label{eq:power1bis}
\end{equation}

\noindent The expression for $P_2$ is obtained simply by swapping the subscripts $1$ and $2$; so the total output power delivered by the two coupled TEGs is given by:

\begin{equation}
P = \alpha_1^2 \alpha_2^2 (T_{\rm hot}-T_{\rm cold})^2 \frac{{R_{\ell}}_1 \alpha_2^2 + {R_{\ell}}_2 \alpha_1^2}{[({R_{\ell}}_1 + R_1) \alpha_2^2 + ({R_{\ell}}_2 + R_2) \alpha_1^2]^2}
\label{eq:powertot}
\end{equation}

\noindent To determine the working conditions for each generator leading to power maximization we solve these two equations:

\begin{equation}
\frac{{\rm d}P}{{\rm d}{R_{\ell}}_1}=0
~~~\mbox{and}~~~
\frac{{\rm d}P}{{\rm d}{R_{\ell}}_2}=0
\end{equation}

\noindent Both lead to the same relation between the electrical resistances and the Seebeck coeffients:

\begin{equation}
\alpha_2^2 ({R_{\ell}}_1 - R_1) = \alpha_1^2 (R_2 - {R_{\ell}}_2),
\label{eq:maximcondition}
\end{equation}

\noindent which reflects the behavior displayed on Fig.~\ref{fig:figure2}.c: power maximization can, in principle, be reached through an infinite number of combinations of working conditions for the two systems. Furthermore, it is in good quantitative agreement with the simulation. It is surprising that satisfaction of electrical impedance matching for both TEGs always leads to a power maximization, as though the TEGs did not experience their mutual thermal influence, in that particular case. Moreover, under the approximations made, the maximum power reads:

\begin{equation}
P_{\rm max} =  \frac{\alpha_1^2 \alpha_2^2 (T_{\rm hot}-T_{\rm cold})^2}{4(R_1 \alpha_2^2 + R_2 \alpha_1^2)},
\label{eq:Pmax}
\end{equation}

\noindent which shows that it is independent of the particular values of the load resistances as long as Eq.~(\ref{eq:maximcondition}) is satisfied.

Now it is instructive to examines the extreme cases when either ${R_{\ell}}_1$ or ${R_{\ell}}_2$ tends to zero: if one of the generators (for exemple TEG$_1$) is electrically shorted (i.e. ${R_{\ell}}_{1} \ll R_1$), it can be viewed as a simple thermal resistance of conductance $K_{\rm contact} = {K_{\rm TEG}}_1 = \alpha_1^2 \overline{T}/R_1$. In that case, and still with the strong coupling assumption, electrical impedance matching is given by ${R_{\ell}}_2 / R_2 = 1 + \alpha_2^2 \overline{T} /(R_2 K_{\rm contact})$ (see Eq. (14) of \cite{Apertet2012b}). After simplifications,

\begin{equation}
{R_{\ell}}_2 = R_2 + \frac{\alpha_2^2}{\alpha_1^2} R_1
\label{eq:limitcase}
\end{equation}

\noindent and we see that we recover exactly Eq.~(\ref{eq:maximcondition}). The same conclusion holds if TEG$_2$ is shorted instead of TEG$_1$; only the subscripts need to be swapped.

\subsection{Efficiency at maximum power}
Let us now turn to the efficiency at maximum power $\eta(P_{\rm max})$, which is given by the ratio of the maximum power $P_{\rm max}$ to the incoming heat flux $I_Q^{\rm (in)}$ that also depends on $P_{\rm max}$:

\begin{equation}
\eta(P_{\rm max}) = \frac{P_{\rm max}}{I_Q^{\rm (in)}(P_{\rm max})}
\label{eq:effPmaxgene}
\end{equation}

\noindent with

\begin{eqnarray}
\nonumber
I_Q^{\rm (in)}(P_{\rm max}) & = &\alpha_1 T_{\rm hot} I_1(P_{\rm max})\\
& = & \alpha_1 T_{\rm hot} \frac{\alpha_1 \left[T_{\rm hot}-T_{\rm m}(P_{\rm max})\right]}{{R_{\ell}}_1 + R_1}
\label{eq:IQinPmax}
\end{eqnarray}

\noindent In the above expression, we see that the dependence of the incoming heat flux $I_Q^{\rm (in)}$ on $P_{\rm max}$ is through the temperature $T_{\rm m}$. To determine $T_{\rm m}(P_{\rm max})$, we use the analytic form of $T_{\rm m}$ given by Eq.~(\ref{eq:Tm}), and the condition for electrical impedance matching, Eq.~(\ref{eq:maximcondition}). Then a straightforward calculation leads to the following simple expression of $\eta(P_{\rm max})$:

\begin{equation}
\eta(P_{\rm max})=\frac{\eta_{\rm C}}{2}
\label{eq:effmaxdemicarnot}
\end{equation}

\noindent We thus recover a well known result, except that this is not through an expansion of $\eta(P_{\rm max})$. In this section we have simplified the problem to fully grasp the subtleties of efficiency at maximum power; in the next section we base our analysis on the results of the simulation shown above in Sec.~(\ref{sec:simu}).

\section{\label{sec:EMP}Analysis of the efficiency at maximum power}

To study the efficiency at maximum power we first have to determine the various working conditions corresponding to this particular engine operation mode. This amounts to locating the coordinates of the maxima of the curves displayed on Fig. \ref{fig:figure2}. We consider the maxima along the lines determined by a constant ratio of ${R_{\ell}}_2/R_2$ to ${R_{\ell}}_1/R_1$. Variation of this ratio yields a description of the whole set of possible combinations of working conditions leading to power maximization [i.e. the maxima of Fig.~(\ref{fig:figure2})].

\begin{figure}
	\centering
		\includegraphics[width=0.5\textwidth]{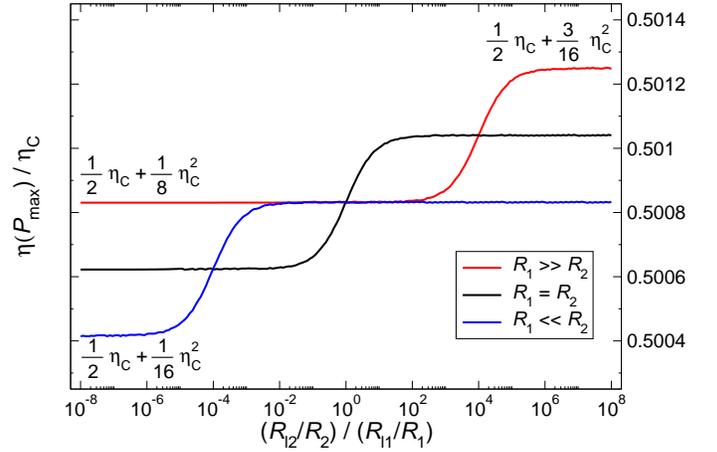}
\caption{Reduced efficiency at maximum power depending on the ratio of ${R_{\ell}}_2$ and ${R_{\ell}}_1$ normalized by $R_1/R_2$. The three curves correspond to the cases where $R_1=R_2$, $R_1 = 10000 \times R_2$ and $R_2 = 10000 \times R_1$.}
	\label{fig:figure4}
\end{figure}

The efficiency associated with the maximum power condition is shown on Fig.~\ref{fig:figure4} as a function of the ratio of ${R_{\ell}}_2/R_2$ to ${R_{\ell}}_1/R_1$, i.e. for various combinations of thermoelectric generators electrical load conditions. We consider three cases: $i/$ the TEGs are identical; $ii/$ the internal resistance of TEG$_1$ is much larger than that of TEG$_2$; $iii/$ the internal resistance of TEG$_1$ is much smaller than that of TEG$_2$. In all cases, the Seebeck coefficients of both generators are equal.

As expected (see e.g. \cite{Apertet2011c, Esposito2009}) $\eta(P_{\rm max})$ is a little larger than $\eta_{\rm C}/2$, and thus different. We find an additional dependence of this efficiency at maximum power on the ratio of internal resistances $R_2/R_1$. We can make a general statement: a stronger dissipation on the hotter side than on the colder side yields a higher efficiency at maximum power. This result is consistent with the case of a single TEG with dissipative thermal contacts \cite{Apertet2011c}.

The condition $({R_{\ell}}_2/R_2) \gg ({R_{\ell}}_1/R_1)$ means that the first TEG is closer to the short-circuit state than the second one and hence that TEG$_1$ dissipates more power than TEG$_2$. Furthermore, if $R_1 \gg R_2$, the capacity of TEG$_1$ to internaly dissipate power is greater than that of TEG$_2$, which results in an increase of $\eta(P_{\rm max})$ : the curve of $\eta(P_{\rm max})$ as a function of $({R_{\ell}}_2/R_2) / ({R_{\ell}}_1/R_1)$ is shifted to higher values.

We note that for the symmetric case ($R_1 = R_2$), the EMP $\eta(P_{\rm max})$ varies around the value found for the case of a single TEG with dissipative contacts to the temperature reservoirs: $\eta_{\rm C}/2 + \eta_{\rm C}^2/8$. The factor $1/8$ at the second order of the expansion in $\eta_{\rm C}$ of $\eta(P_{\rm max})$ is considered as universal for symmetric low-dissipation heat engines operating in the strong coupling regime: first intuited by Tu \cite{Tu2008}, this result was demonstrated later by Esposito and co-workers \cite{Esposito2009}.  

When the ratio $R_1/R_2$ is changed, $\eta(P_{\rm max})$ may vary between $\eta_{\rm C}/2 + \eta_{\rm C}^2/16$ and $\eta_{\rm C}/2 + 3\eta_{\rm C}^2/16$ depending on the ratio of dissipation between hot and cold sides. This result brings back to mind the expression for $\eta(P_{\rm max})$ first derived by Schmiedl and Seifert for a stochastic heat engine \cite{Schmiedl2008}:

\begin{equation}
\eta_{\rm SS}(P_{\rm max})=\frac{\eta_{\rm C}}{2 - \gamma \eta_{\rm C}}
\label{eq:effSS}
\end{equation}

\noindent where the parameter $\gamma$ characterizes the distribution of dissipation. The interest of considering an association of two coupled TEGs rather than a single one, as studied in Ref. \cite{Apertet2011c}, is that $\gamma$ may differ from $1/2$, a value imposed in the case of single TEG by the symmetry of the internal disspation mechanism (Joule effect). If each TEG still obeys to Joule effect symmetry, a dissymetry of either the TEGs themselves ($R_1\neq R_2$), or the working conditions of each module (through ${R_{\ell}}_1$ and ${R_{\ell}}_2$) leads to a significant modification of $\gamma$. Combination of TEGs seems a simple way to achieve tuning of the parameter $\gamma$. 

\begin{figure}
	\centering
		\includegraphics[width=0.5\textwidth]{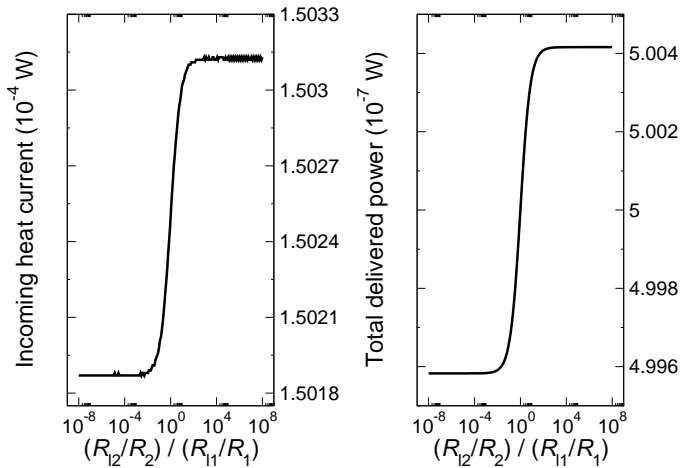}
	\caption{Heat flux from the hot reservoir (left) and maximum power (right) as functions of the ratio of ${R_{\ell}}_2$ to ${R_{\ell}}_1$ normalized by $R_1/R_2$. In this case, $R_1=R_2$.}
	\label{fig:figure5}
\end{figure}

One may wonder about the nature of the dependence of $\eta(P_{\rm max})$ on the ratio $({R_{\ell}}_2/R_2) / ({R_{\ell}}_1/R_1)$. If the maximum power value is actually constant as derived in Sec.~(\ref{sec:analytic}), the increase in efficiency for $({R_{\ell}}_2/R_2) \gg ({R_{\ell}}_1/R_1)$ should be the consequence of the decrease of incoming thermal current in the first TEG. To check this hypothesis we computed both the maximum power value and the associated incoming thermal flux for the case where $R_1=R_2$; the results are plotted in Fig.~(\ref{fig:figure5}). We see that the maximum power and the thermal current actually depend on the specific working conditions of both engines, and that they follow the same trend as $\eta(P_{\rm max})$. This discrepancy with the analytical approach where the maximum power is constant may be explained by the fact that in the analytical derivation we made an approximation for the average temperatures: $(T_{\rm m}+T_{\rm cold})/2 \approx (T_{\rm hot}+T_{\rm m})/2 \approx \overline{T}$; this led to a simplification of $\overline{T}$ in Eq.~(\ref{eq:maximcondition}). We believe that the small variations are, in fact, linked to the ratio of $(T_{\rm m}+T_{\rm cold})/2$ to $(T_{\rm hot}+T_{\rm m})/2$, which is close to one but varies sufficiently with the load resistances to change the conditions for power maximization. To be rigorous, we should say that the true power maximization is reduced to one working condition that the first TEG is shorted, but we considered that the maximum power condition is satisfied for all local maxima. 

We end this section with one surprising fact arising from our analysis of efficiency at maximum power: independently of the specific values of TEGs' parameters ($\alpha_1$, $\alpha_2$, $R_1$ and $R_2$), whenever electrical impedance matching is achieved for both TEGs simultaneously (${R_{\ell}}_1 = R_1$ and ${R_{\ell}}_2 = R_2$) we obtain the value of the Curzon-Ahlborn efficiency at the second order in the Carnot efficiency: $\eta_{\rm C}/2 + \eta_{\rm C}^2/8$, for the EMP of the whole system. For this particular working condition, the two TEGs have respectively the CA efficiency too, which is coherent with Eq.~(\ref{eq:harman}).

\section{\label{sec:intdiss}On the importance of internal dissipation}

Energy dissipation is necessary to ensure the causality of the processes during the operation of heat engines \cite{Apertet2011c}. 
For endoreversible engines as the one presented in Ref.~\cite{Curzon1975}, each flux is related to generalized forces through linear relations, in the engine as well as in the contacts to the reservoirs. Inclusion of internal dissipation in the model obviously breaks the assumption of endoreversibility and necessary implies addition of the dissipated energy in the heat balance. Since the mathematical expression for the quantity of dissipated heat is given by the product of a generalized force by a flux, the additional term corresponding to internal irreversibilities is a quadratic term simply because of the proportionnality between force and flux. Therefore, even though the analysis of the exoreversible engines may have to be performed within the framework of \emph{linear} irreversible thermodynamics, the quadratic term cannot be neglected since it is necessary to ensure the causality of the processes associated to the operation of the engines. Therefore, though the dissipative term is not \emph{stricto sensu} a linear form, we propose to consider it as part of the \emph{linear} irreversible thermodynamics nonetheless, as a direct consequence of the linearity of the constitutive equations of the model. A similar viewpoint was already retained by Callen and Welton in Ref.~\cite{Callen1951}. Recent publications \cite{Izumida2012, Wang2012} addressed this issue of linearity in irreversible thermodynamics. In particular Izumida and Okuda laid the foundations of an extended Onsager model considering the supplementary internal dissipative term \cite{Izumida2012}.  

The following discussion on the importance of the internal dissipation provides further insight into the impact of dissipation localization on EMP. In Ref.~\cite{Apertet2011c} we saw, while studying dissipative thermal contacts, that dissipation occuring on the hot side of the engine enhances its EMP. The analysis of Fig.~(\ref{fig:figure4}) shows that the same trend is recovered here for two thermally coupled TEGs: as already discussed in Sec.~(\ref{sec:EMP}) a dissipation located preferentially on the hot side, with for example $R_1 > R_2$, allows the system's EMP to overcome the Curzon-Ahlborn efficiency bound (or more precisely, the Schmiedl-Seifert efficiency in that case). Note that the idea of heat internally dissipated running back to the hot reservoir, hence allowing better performance of the heat engine was recently discussed in a paper on the thermodynamics of hurricanes \cite{Denur2011}. The author stated that these meteorologic phenomena possess the capacity to preferentially dissipate energy in the higher temperature areas, which in turn enhances their efficiency. Though we will not comment on the derivations made to characterize the hurricanes, this idea of efficiency enhancement due to energy dissipation in the hotter region of a thermodynamic system is consistent with our findings on the behavior of a chain of TEGs.

\begin{figure}
		\includegraphics[width=0.5\textwidth]{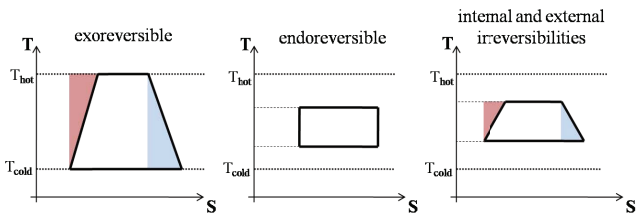}
		\centering
	\caption{Temperature vs. Entropy for different engine models}
	\label{fig:figure7}
\end{figure}

To extend our discussion we consider an irreversible Carnot engine. As this engine experiences friction when the piston moves, energy is dissipated as heat is produced. However, as the chamber is supposed to be adiabatic, heat remains confined inside and may only transit to the reservoirs during the isothermal steps of the cycle. The heat produced during the adiabatic step from $T_{\rm cold}$ to $T_{\rm hot}$ may thus be used during the next cycle. This is analogous to the Joule heating flowing back to the hot reservoir, thereby reducing the incoming heat flow. Adiabatic expansion actually is the key point to understand the importance of the internal disspation inside an exoreversible Carnot engine.

A temperature-entropy ($T-S$) diagram highlights the fact that the adiabatic expansions are not always isentropic for a non-ideal Carnot engines as shown in Ref. \cite{Wu1992}. The behavior of three different configurations for the non-ideal Carnot engine is depicted on Fig.~(\ref{fig:figure7}): on the left side is the exoreversible case where all dissipation takes place inside the engine; in the middle, is the endoreversible case studied by Novikov, and Curzon and Ahlborn; on the right side is a mixed situation involving irreversibilities that occur both internally and externally. The endoreversible cycle is the easiest one of the three to understand: only the temperature of the isothermal steps are modified due to the presence of dissipative thermal contacts. For the exoreversible engine, isothermal steps occur at $T_{\rm hot}$ and $T_{\rm cold}$ respectively; the effect of non-idealities lies in the adiabatic steps: as frictions produce heat, the entropy increases during these steps, but because of the  hypothesis of adiabaticity this heat \emph{remains} in the chamber, the main consequence being that the additionnal heat has to be released into the heat reservoirs during the isothermal steps. This additionnal heat is represented on Fig.~(\ref{fig:figure7}) by the purple triangular area for the step from cold to hot, and by the blue triangular area the step from hot to cold. However these two quantities have totally different impact on the engine performances: the heat produced during the cold to hot step contributes to reduce the absorbed heat during the isothermal step at $T_{\rm hot}$ and hence allows to increase the efficiency. On the contrary, the heat produced during the hot to cold step is released at the cold side and cannot be \emph{recycled}: it is only loss.

Now, considering both internal and external dissipations, we account both for effective temperature relaxation and heat production during the adiabatic steps. In Ref.~\cite{Wu1992}, Wu and Kiang considered this non trivial case. They introduced a quantity $R$ to take account of the ratio of \emph{useful} heat production in cold to hot step and of \emph{pure loss} heat production in hot to cold step. This quantity possesses similarities in its construction with the parameter $\gamma$ appearing in the formula derived by Schmiedl and Seifert, Eq.~(\ref{eq:effSS}). Unfortunately the ratio $R$ as defined in Ref.~\cite{Wu1992} turns out to be innacurate as Wu and Kiang neglected the temperature modification at the engine edges: they considered that the temperature of the isothermal steps remains the same as that obtained for an endoreversible engine, but the heat produced during adiabatic steps actually contributes to a slight increase of these temperatures when frictions are included. This temperature relaxation is similar to the mean temperature modification of a TEG when shifting from endoreversible to exoreversible conditions as illustrated in \cite{Apertet2011c}.

Despite its crucial importance, the adiabatic step often is neglected in the analysis of heat engines. Furthermore, we want to emphasize the fact that Novikov had already considered the fact that internal dissipation plays an important role in the behavior of heat engines. In Ref.~\cite{Novikov1957}, he introduced a parameter $A$ (quite similar to the parameter $\gamma$ of Schmiedl and Seifert, and the ratio $R$ in Wu's and Kiang's work) which characterizes the distribution of the entropy produced by internal irreversibilities between the two adiabatic steps. However in his article, Novikov chose to consider only an endoreversible model and consequently neglected the influence of the parameter $A$.
  
The EMP bound as defined by Gaveau and co-workers \cite{Gaveau2010} is reached for the extremal values of $R$: in the less favorable case, all irreversibilities occur during the hot-to-cold stage, and the EMP is only $\eta^- = \eta_{\rm C} / 2$; whereas in the most favorable case, when all irreversibilities occur during the cold-to-hot stage, the EMP may reach $\eta^+ = \eta_{\rm C} /(2 - \eta_{\rm C})$. In the case of a single TEG, as heat is produced by the Joule effect, it is equally distributed between hot and cold heat reservoirs, thus leaving no degree of freedom to tune significantly the EMP: one can only shift from endoreversible to exoreversible conditions. Nevertheless, for a tandem setup as the one considered in the present article, we can break this natural symmetry by changing both individual working conditions and TEGs parameters. These tuning parameters indeed lead to EMP modifications as shown in Fig.~(\ref{fig:figure4}). However, the bounds $\eta^-$ and $\eta^+$ cannot be reached. Since a thermogenerator must produce Joule heating in order to deliver power, it is impossible to fully release the produced heat to a single heat reservoir; this explains why, in our case, the EMP is comprised between $\eta_{\rm C} / 2 + \eta_{\rm C}^2 / 16$ and $\eta_{\rm C} / 2 + 3 \eta_{\rm C}^2/16$.

\section{\label{sec:discussion}Discussion and conclusive remarks}

\paragraph*{Our viewpoint}
For a tandem construction such as the one proposed in Ref.~\cite{VandenBroeck2005}, there exist several options to tune the working condition of the whole system for given heat reservoir temperatures. Indeed the junction temperature $T_{\rm m}$ and the working conditions for each engine composing the tandem, form three different leverages that can be used for such purpose. However, one should keep in mind that these cannot be employed all at the same time: internal laws governing the system, e.g. heat flux continuity at the interface between engines, determine the third parameter when the other two are fixed. For pratical applications, imposing the electrical load resistances seems the simpler one. Therefore, our point of view is that the temperature profile should not be fixed, but rather deduced from the working conditions of each component of the chain; this view differs from those given in Refs.~\cite{VandenBroeck2005}, \cite{Jimenez2007} and \cite{Jimenez2008}. Interestingly, we recover the fact that, as shown by Van den Broeck \cite{VandenBroeck2005}, when both TEGs work with the Curzon-Ahlborn efficiency, the whole system is working at maximum power with Curzon-Ahlborn efficiency too. However, we have demonstrated that in that case, surprisingly, none of the TEGs works at maximum power.

\paragraph*{Difference with compatibility approach}
In the compatibility approach of thermoelectricity \cite{Clingman1961, Snyder2003, Goupil2009}, the problem is quite different as the TEGs are not only thermally in series but are electrically coupled too. Thus, one cannot tune the working conditions of the TEGs independently: we loose a degree of freedom as only one load resistance, common to the two TEGs, remains, as it is the case in Ref.~\cite{Harman1958}.

\paragraph*{\label{sec:conclusion}Conclusion}
Using a particular tandem construction made of two thermoelectric generators thermally in series but electrically independent, we have derived general results concerning heat engines with a special focus on the efficiency at maximum power. First, we have obtained the working conditions yielding power maximization of the whole system: we noted that this condition is not unique and that it can be obtained for various combinations of the electrical load resistances of each generator. Through analytical derivations, we obtained an expression that defines the couples of resistances maximizing the power. Yet, the analytical approach was not sufficient to fully describe the variations of efficiency at maximum power; so we turned to numerical simulations. We recovered the fact that the Curzon Ahlborn efficiency can be overcome for some specific cases. We then explained the variations of EMP focusing on the importance of internal dissipation. We demonstrated that the distribution of this internal dissipation is the key point to understand issues related to the EMP: if this dissipated heat flows back to the hot heat reservoir, it can be viewed as a recycling process, i.e. one has not to supply this heat during the next isothermal process, whereas dissipated heat that flows to the cold heat reservoir, amounts to pure loss. The tandem construction allowed us to vary this distribution and demonstrate its impact on the efficiency at maximum power.   

\begin{acknowledgments}
This work is part of the CERES 2, ISIS and SYSPACTE projects funded by the Agence Nationale de la Recherche. Y. A. acknowledges financial support from the Minist\`ere de l'Enseignement Sup\'erieur et de la Recherche.
\end{acknowledgments}

\appendix

\section{\label{sec:annexeA}Thermoelectric module operating in the strong flux coupling regime}

Adopting a local description of the operation of a thermoelectric module, the coupled transport of electric charges and heat may be expressed as follows in the force-flux formalism:

\begin{equation}\label{frcflxmicro}
\left(
\begin{array}{c}
\vec{J}\\
~\\
\vec{J_Q}\\
\end{array}
\right)
=
\left(
\begin{array}{cc}
L_{11}~ & ~L_{12}\\
~\\~
L_{21}~ & ~L_{22}\\
\end{array}
\right)
\left(
\begin{array}{c}
\overrightarrow{\nabla}\left(-\frac{\displaystyle \mu}{\displaystyle T}\right)\\
~\\
\overrightarrow{\nabla}\left(\frac{\displaystyle 1}{\displaystyle T}\right)\\
\end{array}
\right),
\end{equation}
where $L_{ij}$ are the kinetic coefficients associated with the phenomenon.

It is often convenient to rewrite the fluxes as a function of the gradients of temperature $T$ and electrochemical potential $\mu$:

\begin{equation}\label{frcflxthermo}
\left(
\begin{array}{c}
\vec{J}\\
~\\
\vec{J_Q}\\
\end{array}
\right)
=
\left(
\begin{array}{cc}
\sigma~ & ~\sigma \alpha\\
~\\~
\sigma \alpha \overline{T}~ & ~\sigma \alpha^2 \overline{T} + \kappa_{_{\vec{J} = \vec{0}}}\\
\end{array}
\right)
\left(
\begin{array}{c}
\overrightarrow{\nabla}\left(-\frac{\displaystyle \mu}{\displaystyle e}\right)\\
~\\
-\overrightarrow{\nabla} T\\
\end{array}
\right),
\end{equation}

\noindent where $\alpha$ is the Seebeck coefficient, $\sigma$ the electrical conductivity, $\kappa_{_{\vec{J}=\vec{0}}}$ the thermal conductivity under open circuit condition and $e$ the elementary electric charge. To evaluate the coupling strength between the electrical and thermal fluxes we introduce the parameter $q$ defined as:

\begin{equation}
q=\frac{L_{12}}{\sqrt{L_{11} L_{22}}}
\end{equation}

\noindent After substitution of the kinetic coefficients by their expressions in the particular case of thermoelectric transport (see Ref.~\cite{Goupil2011}) we obtain:

\begin{equation}
q=\frac{{\rm sgn}(\alpha)}{\sqrt{1 + \frac{\displaystyle \kappa_{{\vec{J}=\vec{0}}}}{\displaystyle \sigma T \alpha^2}}}=\frac{\displaystyle {\rm sgn}(\alpha)}{\displaystyle \sqrt{1 + \frac{\displaystyle 1}{\displaystyle ZT}}},
\end{equation}

\noindent where ${\rm sgn}$ denotes the signum function. At first sight, a large value of the figure of merit $ZT$ is sufficient to ensure a strong flux coupling inside the TEG. However, we should keep in mind that the strong coupling also is characterized by a relation of proportionnality between electrical and thermal fluxes: even the limit $ZT \rightarrow \infty$ is insufficient on its own. The strong coupling regime may be obtained with $\kappa_{{\vec{J}=\vec{0}}}=0$; in other cases there still is an additionnal term in the expression of the thermal current: $\vec{J_Q}=\alpha T \vec{J} - \kappa_{{\vec{J}=\vec{0}}} \vec{\nabla} T$. Furthermore, assuming $\alpha \rightarrow \infty$ necessarily implies that $\vec{J} =\vec{0}$ since it is the only way to avoid a divergence of the thermal flux.

\section{\label{sec:annexeB}Case with no strong flux coupling assumption}
 
\begin{figure}
		\includegraphics[width=0.5\textwidth]{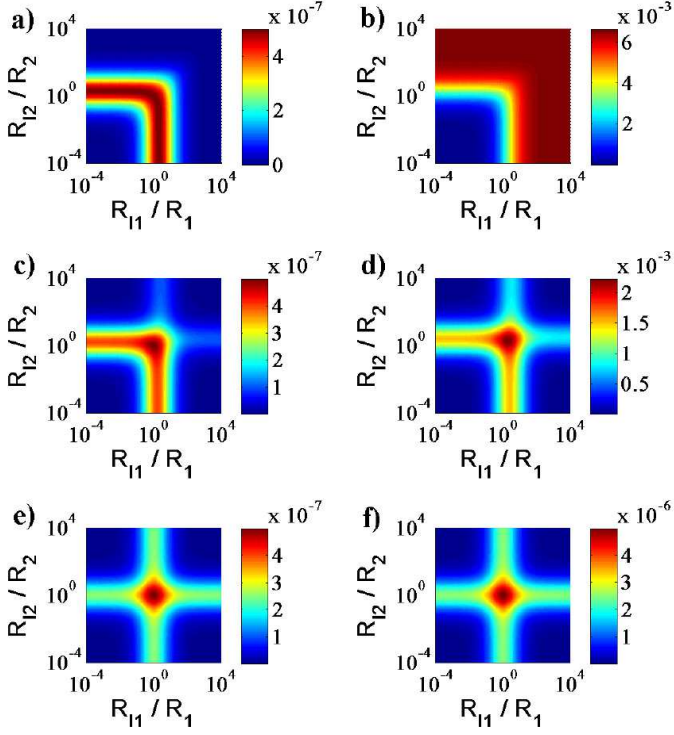}
		\centering
	\caption{Power (first column) and efficiency (second column) for different values of $|q|$ (or equivalently $ZT$). a-b: $|q|$=1 ($ZT = \infty$), c-d: $|q| = 0.87$ ($ZT = 3$), e-f: $|q| = 0.05$ ($ZT = 0.003$).}
	\label{fig:figure6}
\end{figure}

To analyze the influence of the coupling parameter $q$ between the electrical and thermal flux, we choose to tune it keeping constant internal resistance and Seebeck coefficient and only changing the value of the thermal conductance under open-circuit condition $K_{_{I=0}}$. For sake of simplicity, the two generator are always identical. The impact of the $q$ parameter is evidenced in Fig.~(\ref{fig:figure6}). For strong coupling, i.e. $q=1$, the influence of each load resistance is strongly correlated to the working condition of the other TEG. When this hypothesis is relaxed and termal leaks are raised, i.e. $K_{_{I=0}}$ increased, the two TEGs working conditions become more and more independent. When the advective part of the thermal flux, controled by electrical load trough the value of the current, becomes negligeable compared to the conductive heat flow the performance of the whole system are given only by the addition of the two TEGs' taken independently: there is no more correlation between the two subsystems as shown in the third line of Fig.~(\ref{fig:figure6}).

\section{\label{sec:annexeC}Comments on Joule heating}

To satisfy the steady-state heat equation, sometimes referred to Domenicali's equation \cite{Domenicali1954,Domenicali1954b}, one needs to add in an \emph{ad hoc} fashion the heat generated by the Joule effect to the thermal flux balance. Solving $\overrightarrow{\nabla}\cdot\vec{J_Q}= (\overrightarrow{\nabla}\mu)\cdot\vec{J}$, we obtain:

\begin{equation}
-\overrightarrow{\nabla}\cdot\left[\kappa_{_{{\overrightarrow{J}=\vec{0}}}} \overrightarrow{\nabla}T\right] = \frac{{\vec{J}}^2}{\sigma} - T \vec{J}\cdot\overrightarrow{\nabla}\alpha \\
\label{eq:heatequation}
\end{equation}

\noindent Assuming constant system parameters ($\overrightarrow{\nabla}\alpha=\vec{0}$ and $\overrightarrow{\nabla}\kappa_{_{\vec{J}=\vec{0}}}=\vec{0}$), and a constant temperature gradient (given by the ratio of the temperature difference to the length of the module) as in the Onsager-Callen formalism \cite{Onsager1,*Onsager2,Callen1}, the left hand side of Eq.~\eqref{eq:heatequation} reduces to zero. This yields $\vec{J}^2/\sigma=\vec{0}$, which is impossible in the presence of a finite electrical current in the system. To satisfy the heat equation, an extra term must be added to the heat flux.

The force-flux formalism only deals with average quantities inside the heat engine. To account for the difference between input and output fluxes the thermal flux may be made asymmetric by addition of algebraic terms characterizing the total Joule heating, in the incoming and outgoing fluxes. Each end of the system is allocated half the heat as demonstrated in Ref.~\cite{Kawai1994}. This equipartition remains valid at the mesoscopic scale \cite{Gurevich1997}. One way to avoid the \emph{ad hoc} addition of Joule heating is to allow the temperature gradients on the system's input and output sides to be different: to do so one may define an internal temperature. By solving numerically both Onsager-Callen relations and the heat equation, it is easy to show that one does not actually need addition of Joule terms: the system's internal temperature reflects the Joule heating \cite{Apertet2012d}.

\end{document}